\documentclass[aps,10pt,prl,twocolumn,showpacs,nofootinbib,preprintnumbers,superscriptaddress]{revtex4-1}

\usepackage[latin1]{inputenc}
\usepackage{graphicx}
\usepackage{amsmath,amsthm,amssymb}
\usepackage{bm}
\usepackage{url}
\usepackage{hyperref}

\newcommand{\be}{\begin{equation}}
\newcommand{\ee}{\end{equation}}
\newcommand{\ba}{\begin{eqnarray}}
\newcommand{\ea}{\end{eqnarray}}
\newcommand{\V}{{\text{\tiny $V$}}}
\newcommand{\f}{{\text{\tiny f}}}
\def\bs{\begin{subequations}}
\def\es{\end{subequations}}
\def\a{\alpha}

\def\k{\kappa}
\def\e{\epsilon}

\def\s{\sigma}

\def\vp{\varphi}
\def\dpl{\delta_{\rm Pl}}

\newcommand{\Eq}[1]{(\ref{#1})}
\def\lp{\ell_{\rm Pl}}

\def\rmd{d}

\usepackage{color}

\begin{document}

\title{Observational constraints on loop quantum cosmology}

\author{Martin Bojowald}
\affiliation{Institute for Gravitation and the Cosmos,
The Pennsylvania State University,
104 Davey Lab, University Park, Pennsylvania 16802, USA}
\author{Gianluca Calcagni}
\affiliation{Max Planck Institute for Gravitational Physics 
(Albert Einstein Institute), Am M\"uhlenberg 1, D-14476 Golm, Germany}
\author{Shinji Tsujikawa}
\affiliation{Department of Physics, Faculty of Science, 
Tokyo University of Science, 
1-3, Kagurazaka, Shinjuku-ku, Tokyo 162-8601, Japan}


\begin{abstract}
In the inflationary scenario of loop quantum cosmology in the
presence of inverse-volume corrections, we give analytic formulas for
the power spectra of scalar and tensor perturbations 
convenient to compare with observations. 
Since inverse-volume corrections can provide strong
contributions to the running spectral indices, inclusion of terms
higher than the second-order runnings in the power spectra is
crucially important. Using the recent data of cosmic microwave
background and other cosmological experiments, we
place bounds on the quantum corrections.
\end{abstract}

\pacs{98.80.Cq, 04.60.Pp}

\date{January 27, 2011}

\preprint{\href{http://link.aps.org/doi/10.1103/PhysRevLett.107.211302}{Phys.\ Rev.\ Lett.\ {\bf 107}, 211302 (2011)} \qquad [arXiv:1101.5391]}
\maketitle


One of the motivations to search for a quantum theory of gravity is
the desire to unify general relativity with quantum mechanics and,
thereby, resolve classical singularities such as the big bang or those
associated with black holes.
Observational implications of quantum gravity, however, 
present a delicate issue. Based on dimensional grounds, 
cosmology in a nearly isotropic setting
seems to allow quantum corrections only as powers of 
the small quantity $\lp H\approx
10^{-10}$, where $\lp$ is the Planck length and $H^{-1}=a/\dot{a}$ 
is the Hubble radius ($a$ is the scale factor in the flat
Friedmann-Robertson-Walker background and dots denote derivatives with respect
to cosmic time $t$). This dimensional argument is supported by low-energy
effective actions of higher-curvature type.

Dimensional arguments, generally, are overcome if there are more than two
dynamical scales of the same dimension. Detailed physics rather than rough
estimates are then required to determine which geometric mean of the scales is
relevant in a given regime. In cosmology, an additional distance scale $L$
would allow a multitude of dimensionless combinations 
$\lp^{\alpha} H^{\beta} L^{\gamma}$ with
$\alpha-\beta+\gamma=0$, not all of them small. 
Quantum gravity provides ample
motivation for the existence of a third scale by suggesting discrete spatial
structures. While the discreteness scale $L$ is often expected to be near
$\lp$, it is not identical to it and also depends on excitation levels of
states (rather than just Newton's and Planck's constants).

One explicit formulation of such a discrete version of gravity is loop
quantum gravity (LQG) \cite{Revs}. Discreteness arises on the space of metrics
(geometrical operators acquiring discrete spectra). In a nearly homogeneous
quantum space-time, one can think of any region of volume $V$ to consist of
discrete patches, each roughly of size $L^3$ with the length $L$ determined by
an underlying quantum-gravity state. Discrete spectra imply that
derivatives by $L$, as they ubiquitously appear in canonical expressions via
Poisson brackets, are replaced by finite difference quotients. As a simple
example for so-called inverse-volume corrections, $(2\sqrt{L})^{-1}=
d\sqrt{L}/dL$ would, when evaluated for discrete operators, become $(\sqrt{L+\lp}-\sqrt{L-\lp})/2\lp$, which strongly
differs from $(2\sqrt{L})^{-1}$ for $L\sim\lp$. For larger $L$, corrections
are perturbative and of the order $\lp/L$; no factor of $H$ appears. The ratio
$\lp/L$ can easily be much larger than $\lp H$, explaining why this type of
discreteness could give rise to stronger quantum effects.

The results of detailed constructions in LQG, following
\cite{InvScale,InflObs}, will be summarized momentarily. First, we emphasize
that the discreteness does not break general covariance in the equations used
here (assuming small corrections). This has been demonstrated by an
elaborate analysis of the gauge contents of the quantum-corrected theory,
verifying the existence of a closed algebra of gauge generators
\cite{ConstraintAlgebra}. Covariance, and the space-time structure it belongs
to, is then not destroyed but deformed. (Deformations of classical symmetries
play an important role in several approaches to quantum-gravity phenomenology
\cite{DSR}. The deformations considered here are on a different footing,
however, because they do not refer just to Poincar\'e transformations of
Minkowski space.)

Here, using currently available data, we place constraints on
inverse-volume corrections for inflation.
Since scalar and tensor perturbations are subject to
strong modifications of the power on large scales, the corrections are
bounded from above. A detection of gravitational waves and
the precise measurement of cosmic microwave background (CMB) anisotropies in future observations
such as Planck will potentially allow us to make a decisive test for
loop quantum cosmology (LQC) inflation.

A simplified implementation of corrections expected from LQG in
cosmological scenarios via perturbations around homogeneous or other
reduced models can be achieved in LQC \cite{LivRev}.  With a
phenomenological approach to effective dynamics, the cosmological
equations can be summarized in a single Mukhanov equation for the
gauge-invariant scalar perturbation $u_k$, $u_k''+(s^2k^2-z''/z)u_k
=0$ \cite{InflObs} in momentum space with the comoving wave number $k$,
where primes denote derivatives with respect to conformal time
$\tau=\int a^{-1}\,dt$. 
Similarly, tensor modes are subject to the equation
$w_k''+(\a^2k^2-\tilde a''/\tilde a)w_k=0$ \cite{tensor}. Here,
$z(a,\vp)$ and $\tilde{a}(a)$ are background functions and
$\a^2\approx 1+2\a_0\dpl$ and $s^2 = 1+\chi \dpl$ are the propagation
speeds squared, differing from the speed of light by quantum
corrections.

The quantum corrections are characterized by (i)
numerical coefficients $\a_0$ and $\chi$ and (ii) the function
$\dpl\propto a^{-\s}$ determining the size of inverse-volume
corrections.  The values of $\a_0$, $\chi$ and
$\sigma$ are currently subject to quantization ambiguities. $\chi$ is parametrized as
$\chi=\sigma\nu_0(\sigma/6+1)/3+\a_0(5-\sigma/3)/2$,
where $\nu_0$ is related to $\a_0$ and $\sigma$
by the consistency condition \cite{InflObs}
\be
\nu_0(\sigma-3)=3 \alpha_0 (\sigma-6)/(\sigma+6)\,.
\label{concon}
\ee While $\sigma$ takes values in the range $0<\sigma\leq 6$, the
size of $\dpl$ does not depend on the values of $\a_0$ and $\nu_0$.
With $\sigma>0$, $\dpl$ is larger at early times, in agreement with
discreteness departing from the Planck scale in a more classical
universe. The aim of this Letter is to restrict $\dpl$ by observations.
We will mainly place bounds on the combination
$\alpha_0\dpl$ during slow-roll (SR) inflation, for which the precise
origin of $\alpha_0$ and $\nu_0$ or the scale hidden in $\dpl$ is not
essential.

Corrections in the evolution equations arise only in the $k^2$ term,
not in the time derivative of the d'Alembertian, yet they are
covariant according to the corrected gauge transformations
\cite{ConstraintAlgebra}.  Thus, one typical assumption of
higher-curvature theories is violated.  Moreover, the propagation
speed of tensor modes differs from the scalar one since in
general $2\a_0\neq\chi$. Again, this is only possible with the change
in the underlying manifold and gauge structure, and gives rise to
additional characteristic effects. With different types of equations
for scalar and tensor modes, there are changes to the standard
inflationary spectra and the tensor-to-scalar ratio.

In Ref.~\cite{InflObs}, two of us evaluated the inflationary
observables in terms of the three SR parameters
$\epsilon=-\dot{H}/H^2$, $\eta=-\ddot{\varphi}/(H \dot{\varphi})$, and
$\xi^2= (\ddot{\varphi}/\dot{\varphi})^{\dot{}}/H^2$, where $\varphi$
is a scalar field with potential $V(\varphi)$.  In order to place
observational bounds on concrete inflaton potentials, it is more
convenient to use SR parameters expressed by $V$ and its derivatives:
$\e_\V\equiv \k^{-2}(V_{,\vp}/V)^2/2$,
$\eta_\V\equiv\k^{-2}V_{,\vp\vp}/V$, $\xi_\V^2 \equiv
\kappa^{-4}V_{,\vp}V_{,\vp \vp \vp}/ V^2$ where $\k^2=8\pi G$ ($G$ is
the gravitational constant).  
For conversion formulas from $\epsilon, \eta, \xi^2$
to $ \e_\V, \eta_\V, \xi_\V^2$ and all the technical details we refer to
\cite{InflConsist}, together with a discussion of cosmic variance.

The power spectra of scalar and tensor perturbations, 
evaluated at the Hubble horizon crossing during inflation 
($k \approx aH$), are given, respectively, by \cite{InflObs} 
\be
{\cal P}_{\rm s}=\frac{GH^2}{\pi \epsilon}
(1+\gamma_{\rm s} \dpl)\,,\quad
{\cal P}_{\rm t}=\frac{16 GH^2}{\pi} 
(1+\gamma_{\rm t} \dpl )\,,
\label{powerori}
\ee
where 
$\gamma_{\rm s}=\nu_0 (\sigma/6+1)+\sigma \alpha_0/(2\epsilon)
-[\sigma \nu_0 (\sigma+6)+3\alpha_0 (15-\sigma)]/[18 (\sigma+1)]$
and $\gamma_{\rm t}=(\sigma-1)\alpha_0/(\sigma+1)$.
We expand the scalar spectrum about a pivot wave number $k_0$, as 
\ba
\ln {\cal P}_{\rm s} (k) &=& \ln {\cal P}_{\rm s} (k_0)
+\left[n_{\rm s} (k_0)-1 \right] x 
\nonumber \\
& &+\frac{\alpha_{\rm s} (k_0)}{2}x^2+\sum_{m=3}^{\infty} 
\frac{\alpha_{\rm s}^{(m)}(k_0)}{m!} x^m\,,
\label{Power}
\ea
where $x=\ln (k/k_0)$, $n_{\rm s}(k)-1\equiv d \ln {\cal P}_{\rm s}(k)/d \ln k$, and 
$\alpha_{\rm s}^{(m)}(k)\equiv d^{m-2} \alpha_{\rm s}/(d\ln k)^{m-2}$.
The tensor spectrum can be expanded in a similar way with 
a different index $n_{\rm t}(k)\equiv d \ln {\cal P}_{\rm t}(k)/d \ln
k$. While such expansions to second order are standard 
in cosmology, terms of order
higher than 2 will become important in our analysis.

The spectral indices are
\be
n_{\rm s}-1 = -6\e_\V+2\eta_\V
-c_{n_{\rm s}}\dpl\,, \quad
n_{\rm t} = -2\e_\V-c_{n_{\rm t}}\dpl\,,
\ee
with quantum-gravity corrections $c_{n_{\rm s,t}}=f_{\rm s,t}+\cdots$ 
whose dominant contributions are
$f_{\rm s} \equiv
\sigma [3\a_0(13\s-3)+ \nu_0\s(6+11\s)]/[18(\sigma+1)]$ and $f_{\rm t}
\equiv 2\sigma^2 \alpha_0/(\sigma+1)$.  For $\sigma\gtrsim O(1)$
the variation of $\dpl$ is fast ($\dpl \propto a^{-\sigma}\propto
k^{-\sigma}$ at Hubble crossing), so that $f_{\rm s,t}$ provide
dominant contributions to the scalar and tensor runnings as well,
$\alpha_{\rm s,t}(k_0)\equiv d n_{\rm s,t}/d \ln k|_{k=k_0}  \approx \sigma
f_{\rm s,t} \dpl (k_0)$.  Similarly, the $m$th order terms are
$\alpha_{\rm s,t}^{(m)} (k_0)  \approx (-1)^m \s^{m-1}f_{\rm s,t}\dpl
(k_0)$ and hence we can evaluate the sum in Eq.~\Eq{Power} as
\be
\sum_{m=3}^{\infty} \frac{\alpha_{\rm s,t}^{(m)}}{m!}x^m
= \left[x\left(1-\frac12 \sigma x \right)
+\frac{e^{-\sigma x}-1}{\sigma}\right]f_{\rm s,t}\dpl.
\label{higher}
\ee
This expression is valid for any value of $\sigma$ and 
of the pivot scale $k_0$ within the observational 
range of CMB. Since the LQC corrections to the runnings 
$\alpha_{\rm s,t}$ can be large, inclusion of the higher-order 
terms (\ref{higher}) is important to estimate the 
power spectra properly. 

For the CMB likelihood analysis we also take into account 
the second-order terms of slow-roll parameters, i.e.
$\alpha_{\rm s}= -24 \e_\V^2+16\e_\V\eta_\V
-2\xi_\V^2+c_{\alpha_{\rm s}}\dpl$ and 
$\alpha_{\rm t}=-4 \e_\V \left( 2\e_\V-\eta_\V \right)
+c_{\alpha_{\rm t}}\dpl$, where the dominant contributions 
to $c_{\alpha_{\rm s,t}}$ correspond to 
$c_{\alpha_{\rm s,t}}\approx\sigma f_{\rm s,t}$. 
In the numerical code, the full expressions of the coefficients 
$c_{n_{\rm s,t}}$ and 
$c_{\alpha_{\rm s,t}}$ \cite{InflConsist} are used.
At the pivot scale $k_0$ we have the tensor-to-scalar ratio 
$r(k_0)\equiv {\cal P}_{\rm t}(k_0)/{\cal P}_{\rm s}(k_0)
= 16\e_\V(k_0)+c_{r}\dpl (k_0)$, 
where $c_{r}=8[3\a_0(3+5\s+6\s^2)-\nu_0\s(6+11\s)]
\e_\V(k_0)/[9(\sigma+1)]
-16\s\a_0 \eta_\V(k_0)/3$.

\begin{figure}
\includegraphics[width=8.0cm]{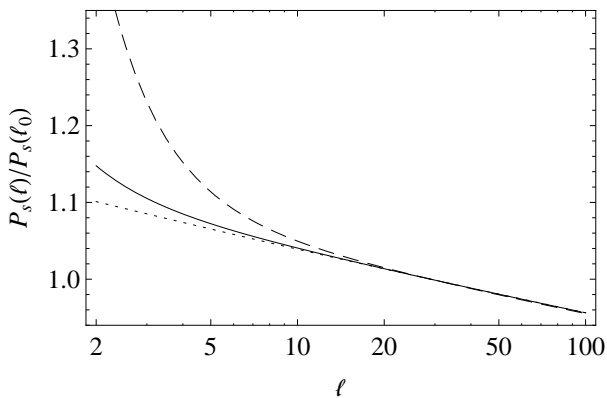}
\caption{Primordial scalar power spectrum ${\cal P}_{\rm s}(\ell)$
for the case $n=2$, $\sigma=2$, and 
$\epsilon_{\V}(k_0)=0.009$ with three different values of $\delta(k_0)$: 0 (classical case, dotted line), $7\times 10^{-5}$ (experimental upper bound, solid line), $4.8\times 10^{-4}$ ($1/10$ of the a-priori
upper bound, dashed line). Here the pivot wave number is $k_0=0.002$\,Mpc$^{-1}$, which corresponds to $\ell_0=29$. 
\label{spectrum}
}
\end{figure}

In the quasi-de Sitter background, $\dpl \propto
k^{-\sigma}$ gives the relation $\dpl (k)  \approx \dpl (k_0)
(k/k_0)^{-\sigma}= \dpl (\ell_0) (\ell/\ell_0)^{-\sigma}$, where
$\ell$ are the CMB multipoles related to $k$ via $k
 \approx (h/10^4)\ell$~{\rm Mpc}$^{-1}$ 
($h \approx 0.7$ is the reduced Hubble constant).  
With the large-volume expansion of quantum
corrections, we require that $\dpl(k) \ll 1$ at all scales.
For $\sigma>0$ the LQC correction is most significant on the largest
scales observed in the CMB ($\ell=2$). This property can be clearly seen
in Fig.~\ref{spectrum}, where the pivot scale for the scalar power
spectrum is taken to be $\ell_0= 29$. Intuitively, this happens because the
largest cosmological scales correspond to those where and when space-time
quantum effects were larger,
while smaller scales have been affected by ordinary physics. Imposing the condition $\dpl
(\ell=2)  \ll 1$, this gives 
\be
\dpl (\ell_0) \ll  (2/\ell_0)^{\sigma}
\label{dplbound}
\ee
at the multipole $\ell_0$. For larger $\sigma$ and $\ell_0$, $\dpl(\ell_0)$ is 
constrained to be smaller. We assume that inflation completely takes place in the large-volume regime, and that longer-wavelength modes that might violate the bound \Eq{dplbound} do not enter the inflationary Fourier analysis.

For concreteness, let us consider the power-law potential
$V(\varphi)=\lambda \varphi^n$, for which $\epsilon_\V
=n^2/(2\kappa^2 \varphi^2)$ and 
\be
\eta_\V=\frac{2(n-1)}{n}\e_\V\,,\qquad
\xi_\V^2 =\frac{4(n-1)(n-2)}{n^2}\e_\V^2\,.
\ee
Among the variables $\sigma$, $\alpha_0$, and $\nu_0$
we have the relation (\ref{concon}), a condition under which,
for given $n$ and $\sigma$, 
the inflationary observables can be expressed via $\epsilon_\V$ and 
$\delta \equiv \alpha_0 \dpl$ for $\s\neq 3$, or by $\e_\V$ and 
$\tilde\delta \equiv \nu_0 \dpl$ for $\s=3$.

We carry out the CMB likelihood analysis by varying the parameters
$\e_\V$ and $\delta$ in the cosmological Monte Carlo (CosmoMC) code
\cite{Antony}.  We use the 7-year WMAP data combined with large-scale structure, the Hubble constant measurement from the Hubble Space Telescope, supernovae type Ia, and 
big bang nucleosynthesis \cite{Komatsu}. 
We assume the flat $\Lambda$-Cold-Dark-Matter 
model with no fraction of massive neutrinos in the dark matter density ($f_{\nu}=0$). 

In the likelihood analysis, we vary the following eight parameters: 
(i) baryon density today, $\Omega_{b}$, 
(ii) dark matter density today, $\Omega_c$,
(iii) the ratio of the sound horizon to the angular 
diameter distance, $\theta$, 
(iv) the reionization optical depth, $\tau$;,
(v) $\delta(k_0)$, (vi) $\e_\V(k_0)$, 
(vii) ${\cal P}_{\rm s}$ ($k_0$), and 
(viii) the Sunyaev-Zel'dovich amplitude, $A_{\rm SZ}$. 
We take the pivot wave number $k_0=0.002$
Mpc$^{-1}$ ($\ell_0\approx 29$) used by the WMAP team. 
$\delta (k_0)$ and $\e_\V (k_0)$ are constrained at this scale. 
While the bound on $\delta$ depends on the pivot scale 
(and it tends to be smaller for larger $k_0$), that on 
$(k_0)^\s\delta(k_0)$ does not.

While we assume a standard treatment of the 
reionization with a smooth interpolation, 
more general reionization 
scenarios can potentially affect constraints on 
observables especially for $n_{\rm s}>1$ \cite{Pand}.
The analysis in \cite{Pand} shows
that the allowed region with $n_{\rm s}<1$ is not strongly modified, 
which is the case for our potentials.

The exponential term $e^{-\sigma x}=(k_0/k)^\sigma$ in Eq.~(\ref{higher})
gives rise to the enhancement of the power spectra on large scales, 
as we see in Fig.~\ref{spectrum}. 
In this sense the LQC corrections can be distinguished from 
the suppression effects coming from the noncommutative geometry 
or string corrections \cite{Maartens}.
For $\sigma\gtrsim 3$, the growth 
of the term $e^{-\sigma x}$ is so significant 
that $\dpl(\ell)$ must be very much 
smaller than 1 for most of the scales observed in the CMB, 
in order to satisfy the bound $\dpl (\ell=2)  \ll 1$. 
More precisely, LQC corrections manifest themselves mainly at $\ell=2,3$, 
where cosmic variance dominates, so it seems implausible to 
isolate these effects. For $\sigma<3$, the LQC modification to 
the classical power spectra also affects larger multipoles $\ell$, 
and hence it is possible to constrain it from CMB anisotropies.

\begin{figure}
\includegraphics[width=7.0cm]{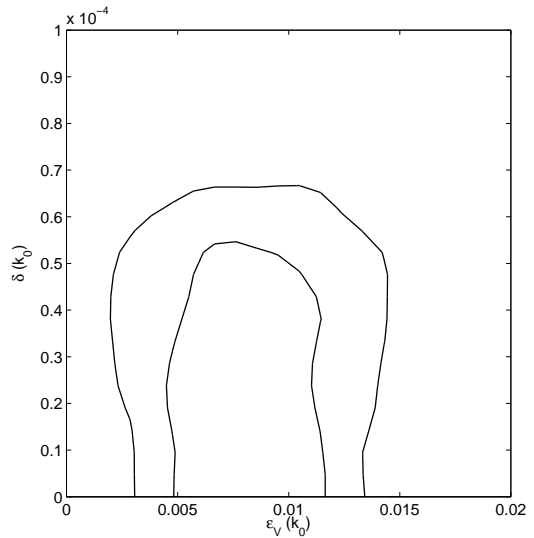}
\caption{2D marginalized distribution for the quantum-gravity parameter
$\delta(k_0)=\a_0\dpl (k_0)$ and the slow-roll parameter 
$\e_\V (k_0)$ with the pivot $k_0=0.002$ Mpc$^{-1}$
for $n=2$ and $\sigma=2$.
The internal and external solid lines correspond to 
the 68\% and 95\% confidence levels, respectively. 
\label{sigma2}}
\end{figure}

\begin{figure}
\includegraphics[width=7.0cm]{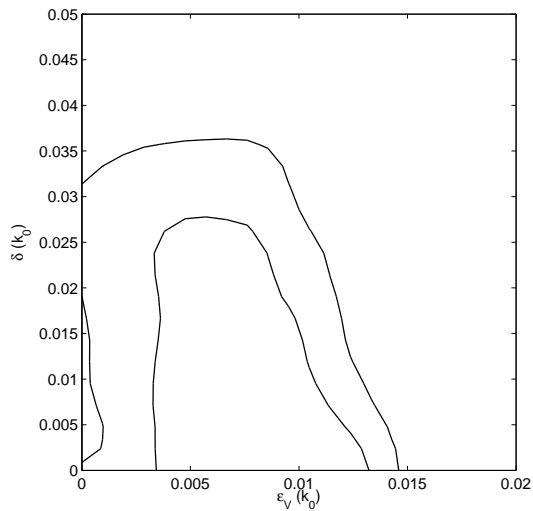}
\caption{
2D marginalized distributions as in  
Fig.~\ref{sigma2}, but for the case 
$n=2$ and $\sigma=1$.
\label{sigma1}}
\end{figure}

In Fig.~\ref{sigma2} we plot the 2D posterior distributions on the
parameters $\delta (k_0)$ and $\epsilon_\V (k_0)$ with $k_0=0.002$
Mpc$^{-1}$ for $n=2$ and $\sigma=2$.  The two parameters are
constrained to be $\delta (k_0)<6.7 \times 10^{-5}$ and
$\epsilon_\V(k_0)<0.013$ (95\% C.L.).  The modification of the
large-scale power spectra ($\ell \lesssim 20$) shown in
Fig.~\ref{spectrum} leads to the upper bound on $\delta (k_0)$.
The condition (\ref{dplbound}) gives the prior 
$\dpl (\ell_0) \ll  4.8 \times 10^{-3}$
at $\ell_0=29$, so that for $\alpha_0=O(1)$
the observational bound is smaller by 2 orders 
of magnitude. For larger $k_0$ the observational 
upper bounds on $\delta (k_0)$ tend to be smaller
for given $\sigma$.
For $k_0=0.05$ Mpc$^{-1}$ and $\sigma=2$, we find that
$\delta (k_0)<1.2 \times 10^{-7}$ (95\% C.L.), in which case
the theoretical expected amplitude [$\delta_{\rm Pl} (k_0) \sim 10^{-8}$ 
or a few orders of magnitude higher \cite{InflObs}] 
can be accessible.

For smaller $\sigma$ the observational upper bound on $\delta (k_0)$
tends to be larger, with milder enhancement of the power spectra on
large scales. In Fig.~\ref{sigma1} we show the likelihood results for
$\sigma=1$, in which case the LQC correction is constrained to be
$\delta (k_0)<3.6 \times 10^{-2}$ (95\% C.L.).
Meanwhile, the a-priori criterion (\ref{dplbound}) gives 
$\dpl(k_0) \ll 6.9 \times 10^{-2}$. For $\alpha_0=O(1)$, the case
$\s=1$ is marginally consistent with the combined SR/$\dpl$
truncation.

For $\sigma \lesssim 1$, the exponential factor $e^{-\sigma x}$ does not 
change rapidly with smaller values of $f_{\rm s,t}$, so that the
LQC effect on the power spectra would not be very significant even if 
$\delta (k_0)$ was as large as $\e_\V (k_0)$.
Our likelihood analysis shows that the observational 
upper bound on $\delta (k_0)$ exceeds the a-priori
upper limit of
$\dpl (k_0)$ given by Eq.~(\ref{dplbound}).
Since $\delta (k_0)$ can be as large as 1, 
the validity of the approximation $\delta (k_0)<\e_\V (k_0)$ 
used in the main formulas may break down in such cases.

Under the conditions $\epsilon_\V, \delta \ll 1$, 
it follows that 
$\epsilon_\V \approx (\kappa^2/2)(\dot{\varphi}/H)^2$.
Then the number of e-foldings during inflation is given by 
$N \equiv \int_{t}^{t_\f} \rmd\tilde{t}\,H \approx \kappa 
\int_{\varphi_\f}^{\varphi} \rmd\tilde{\varphi}/\sqrt{2\epsilon_\V (\tilde{\varphi})}$, 
where $\varphi_{\rm f}$ is the field value at the end of inflation [determined 
by the condition $\epsilon_\V \approx O(1)$].
For the power-law potentials one has $N \approx n/(4\epsilon_\V)-n/4$, 
which gives $\epsilon_\V \approx n/(4N+n)$.
For $n=2$, the theoretically constrained range $45<N<65$ 
corresponds to $0.008<\epsilon_\V<0.011$.
The probability distributions of $\epsilon_\V$ in Figs.~\ref{sigma2} and \ref{sigma1} 
are consistent with this range even in the presence of 
the LQC corrections, 
so the quadratic potential is compatible with observations as in standard cosmology.

In summary, in inflation combined with LQC inverse-volume corrections
we provided general formulas for the scalar and tensor power spectra
and placed observational bounds on the size of corrections for a
quadratic potential. In \cite{InflConsist} we ran the CosmoMC code
also for other potentials such as $V\propto \vp^4$ and $V\propto
e^{-\lambda\vp}$ (for which the inflationary observables reduce,
again, to $\delta$ and $\e_\V$). We found that the observational upper
bounds are practically independent of the inflaton potentials. This is
because the LQC correction is approximately given by $\dpl (k) \approx
\dpl (k_0) (k/k_0)^{-\sigma}$, which only depends on $\s$ and the
pivot scale $k_0$.  Interesting and nontrivial effects do arise from
the modified space-time structure underlying the dynamics. Even though
quantum-geometry corrections are small, they can significantly change
the runnings of spectral indices. Thus, the observational bounds on
$\dpl$ can be much closer to theoretical values
[$O(10^{-8})$] than often thought in quantum gravity.
Our new techniques set the stage for systematic and
stringent phenomenological evaluations.

\smallskip

\noindent M.~B. was supported by NSF Grant No.\ PHY0748336.
S.~T. thanks JSPS for financial support
(Grants No.~30318802 and No.~21111006).



\begin{thebibliography}{99}
{\footnotesize

\bibitem{Revs}
C.~Rovelli, {\em Quantum Gravity} 
(Cambridge University Press, Cambridge, UK, 2004);
A.~Ashtekar and J.~Lewandowski,
Classical Quantum Gravity  {\bf 21}, R53 (2004);
T.~Thiemann, {\em Introduction to Modern Canonical Quantum General Relativity} 
(Cambridge University Press, Cambridge, UK, 2007).

\bibitem{InvScale} 
T.~Thiemann,
Classical Quantum Gravity {\bf 15}, 839 (1998);
%
M.~Bojowald {\it et al.},
Phys.\ Rev.\  D {\bf 75}, 064022 (2007).

\bibitem{InflObs} 
M.~Bojowald and G.~Calcagni, J.\ Cosmol.\ Astropart.\ Phys.\ 03 (2011) 032.

\bibitem{ConstraintAlgebra} 
M.~Bojowald {\it et. al., }
Phys.\ Rev.\  D {\bf 78}, 063547 (2008); 
Phys.\ Rev.\  D {\bf 79}, 043505 (2009).

\bibitem{DSR} 
J.~Magueijo and L.~Smolin, Phys.\ Rev.\ Lett.\ {\bf 88}, 190403 (2002); G.~Amelino-Camelia, Nature (London) {\bf 418}, 34 (2002).


\bibitem{LivRev}
M.~Bojowald, Living Rev.\ Relativity {\bf 11}, 4 (2008), http://www.livingreviews.org/lrr-2008-4.
  
\bibitem{tensor}
M.~Bojowald and G.~M.~Hossain,
Phys.\ Rev.\  D {\bf 77}, 023508 (2008);
G.~Calcagni and G.~M.~Hossain,
Adv.\ Sci.\ Lett.\  {\bf 2}, 184 (2009).

\bibitem{InflConsist} 
M.~Bojowald, G.~Calcagni, and S.~Tsujikawa, arXiv:1107.1540.

\bibitem{Antony} 
http://cosmologist.info/cosmomc/.

\bibitem{Komatsu}
E.~Komatsu {\it et al.}, Astrophys.\ J.\ Suppl.\ {\bf 192}, 18 (2011); B.~A.~Reid {\it et al.}, Mon.\ Not.\ Roy.\ Astron.\ Soc.\ \textbf{404}, 
60 (2010); A.G.~Riess {\it et al.}, Astrophys.\ J.\ \textbf{699}, 539 (2009); M.~Kowalski {\it et al.}, Astrophys.\ J.\ \textbf{686}, 749 (2008); S.~Burles and D.~Tytler, Astrophys.\ J.\ \textbf{499}, 699 (1998).

\bibitem{Pand} 
S.~Pandolfi {\it et al.}, 
Phys.\ Rev.\ D {\bf 82}, 123527 (2010).

\bibitem{Maartens}
S.~Tsujikawa {\it et al.},
Phys.\ Lett.\  B {\bf 574}, 141 (2003);
Y.S.~Piao {\it et al.},
Classical Quantum Gravity  {\bf 21}, 4455 (2004);  G.~Calcagni and S.~Tsujikawa,
  Phys.\ Rev.\ D {\bf 70}, 103514 (2004).
}
\end{thebibliography}
\end{document}